# Combined MEG and fMRI Exponential Random Graph Modeling for inferring functional Brain Connectivity.

Roseric Azondekon[1], Zachary James Harper[1,2], and Charles Michael Welzig[2]

[1]University of Wisconsin Milwaukee, Milwaukee, WI, USA

[2]Medical College of Wisconsin, Milwaukee, WI, USA

**Abstract**

Estimated connectomes by the means of neuroimaging techniques have enriched our knowledge of the organizational properties of the brain leading to the development of network-based clinical diagnostics. Unfortunately, to date, many of those network-based clinical diagnostics tools, based on the mere description of isolated instances of observed connectomes are noisy estimates of the true connectivity network. Modeling brain connectivity networks is therefore important to better explain the functional organization of the brain and allow inference of specific brain properties. In this report, we present pilot results on the modeling of combined MEG and fMRI neuroimaging data acquired during an n-back memory task experiment. We adopted a pooled Exponential Random Graph Model (ERGM) as a network statistical model to capture the underlying process in functional brain networks of 9 subjects' MEG and fMRI data out of 32 during a 0-back vs 2-back memory task experiment. Our results suggested strong evidence that all the functional connectomes of the 9 subjects have small world properties. A group level comparison using a non-parametric paired permutation t-test comparing the conditions pairwise showed no significant difference in the functional connectomes across the subjects. Our pooled ERGMs successfully reproduced important brain properties such as functional segregation and functional integration. However, the ERGMs reproducing the functional segregation of the brain networks discriminated between the 0-back and 2-back conditions while the models reproducing both properties failed to successfully discriminate between both conditions. The pilot results presented here are promising and would improve in robustness with a larger sample size. Nevertheless, our pilot results tend to support previous findings that functional segregation and integration are sufficient to statistically reproduce the main properties of brain network.

# BACKGROUND

The development of sophisticated neuroimaging techniques has enabled the acquisition of non-invasive quantitative data prompting to the development of new concept of the analyses of these data. In the existing literature, estimated connectomes by the means of neuroimaging techniques have enriched our knowledge of the organizational properties of the brain and enabled the development of network-based clinical diagnostics of certain pathologies such as schizophrenia [1], stroke [2], and Alzheimer's disease [3]. Although the mere descriptive analyses of the functional brain connectivity used in those researches have improved our knowledge of brain connectivity maps, there remains a gap in the literature since the description of isolated instances of observed connectivity network are noisy estimates of the true connectivity network [4,5]. In fact, the brain functioning can be represented as a connectivity network (connectome) where parcels or anatomical regions or regions of interest (ROIs) of the brain represent the vertices and the edges determine statistical dependency of combined neuronal activities between the vertices [6].

Modeling brain connectivity networks is therefore important to better explain the functional organization of the brain and to allow inference of specific brain properties. At first, three main mathematical models referred as null models or generative models have been proposed to infer some observed basic network properties such as network size, connection density, and degree distribution. The first is the simple random network model proposed by Erdős and Rényi [7]; a more general formulation of this model was described by Gilbert [8]. Random network models help to hypothesis testing whether the topology of a brain connectivity network arise purely by chance. The second model was proposed by Watts-Strogatz [9] and termed as the Watts-Strogatz small-world model. This model generates random networks spanning at the middle ground of the topological spectrum of random networks and lattice networks. Small world networks are characterized by a relatively high clustering coefficient and a small average path length between nodes. The third model is the preferential attachment model proposed by Barabási and Albert [10]. This model generates more realistic, scale-free degree distribution networks from the concept of "the rich get richer". Although these models allow hypothesis testing and the identification of relevant network properties, they come up short at explaining the organizational mechanisms of brain connectivity network formation [5]. In addition, these mathematical models are not estimable from the observed data, do not allow fitness to the data, and hence cannot provide a reasonable representation of the observed network [11].

To remedy the limitations of generative null models, statistical models have been proposed to not only support inference but to capture and explain the process underlying the formation of the network structure. Unlike mathematical network models, statistical network models are designed to consider all the alternative networks estimated and weighted from observed data [12]. Furthermore, they specifically allow the assessment of significance of terms in the model and evaluation thanks to the goodness of fit. To date, three classes of such statistical network models have been proposed: the class of exponential random graph models, the class of stochastic block models, and the class of latent network models [11]. Analogous to standard regression models, the class of exponential random graph models (ERGM) also referred to as p* models (ERGM family models) appears as a flexible choice to simultaneously assess the role of specific network features in the overall organization of the complexity of brain networks. ERGM based connectivity analyses can help simulate and discriminate normal and abnormal brain organization and functioning [13]. In the social science literature, p* models prove successful at studying complex network interactions [14–18].

In neuroscience, the application of p* family models is still limited as very few studies have proved to successfully use them to model neuroimaging data based connectomes. To the best of our knowledge, the first study of this kind was reported in 2011 by Simpsons et. al [19] who applied ERGM on connectomes derived from 10 fMRI data collected from 10 subjects. Another study conducted in 2016 was reported by Sinke et. al [20] who applied Bayesian ERGM on diffusion tensor imaging (DTI) collected from 382 healthy subjects. More recently, in 2018, Obando and De Vico Fallani [5] published the first study to model functional connectomes derived from EEG data collected on 108 subjects during eyes-open (EO) and eyes-closed (EC) resting-state conditions. While it is understandable that all those studies pioneered the use of p* family models on neuroimaging connectomes, the applicability of ERGM family models to other connectomes inferred from other neuroimaging data is yet to be proved. In this report, we described how we applied p* models to combined MEG and fMRI neuroimaging data acquired during a memory task experiment.

There also remains many methodological unanswered issues such as the connectivity metrics to derive network topology, the ERGM terms to include in the modeling process, as well as how ERGM must be fit to the subject's connectomes. Simpsons et. al [19] and Obando and De Vico Fallani [5] for instance, fit a single ERG model to each subject data. Such a methodological approach lacks robustness when for example, one seeks to estimate a single model that discriminates between EO and EC resting state conditions. ERGM family models have a lot of potentials, especially in providing a better and more robust alternative network-based diagnostic model to the descriptive network-based diagnostic methods of medical conditions [1–3,21,22]. We address the lack of robustness from the previous studies by taking a pooled ERGM approach combining functional connectomes across subjects for each condition.

To the best of our knowledge, this report is the first to ever describe the application of ERGM to combined MEG and fMRI data.

## METHODS

### Participants

Participants were 32 healthy, right-handed adults between the ages of 18 and 40 recruited from the community using local print and electronic media. Recruited participants were all English speakers with at least 12 years of education. No exclusion was made on the basis of race, ethnicity, or gender. Because of the MRI scans, all participants were assessed for contraindications to MRI scanning, such as implanted electronic devices or ferrous metal in sensitive areas.

### Experiment

The participants were asked to perform n-back memory tasks during MEG scans. In our n-back tasks, participants are presented a sequence of visual stimuli one-by-one. For each stimulus, they need to decide if the current stimulus is the same as the one presented n trials ago. Specifically, the participants performed 0-back and 2-back memory tasks during which they are asked to match geometric shapes. The MEG paradigm consists of nine experimental blocks: two blocks each of matching pictures with five control blocks, each lasting 32s for a total scan length of 4:48 min.

Each block begins with a brief instruction statement: "Match Faces" or "Match Tools". Each matching block consists of six images. For each face block, three images of each gender and target affect is presented. All images are presented sequentially, with no inter-stimulus interval, for a period of 5s and in a randomized fashion for both 0-back and 2-back memory tasks. The order of the paradigm is counterbalanced across subjects. During MEG recordings, subjects respond by pressing a button on one of two button boxes, allowing for the determination of accuracy and reaction time.

**MEG acquisition**

All participants undergo MEG scanning at the Medical College of Wisconsin (MCW) MEG lab. Before the experiment, a Polhemus Isotrak® system is used to digitize participants' cardinal landmarks (nasion and pre-auricular points) and head shape. Four head position indicator coils are fixed to the participants' head and referenced to the other digitized landmarks. Two electrodes are placed along the plane of the chest to collect ECG signal. MEG data are acquired with the participant seated upright in the scanner. Data are sampled at 2,000 Hz. The scanning session consists in two to five runs of 10 minutes each. Prior to each subject's scanning session, one to two runs of five to 10 minutes each of empty room MEG data are recorded for noise characterization. In addition, one to two runs of 10 minutes of Eyes-Open (EO) resting state of MEG data are also recorded after the experimental runs. All MEG scanning sessions take place on a different day than MRI scanning sessions.

**MRI acquisition**

All participants undergo high-resolution T1-weighted structural MRI at the MCW 7 Tesla MRI facility. MRI scanning sessions include localizer scans and a GE SPGR T1 acquisition with approximately 1x1x1 mm voxel size and parameters optimized for grey-white contrast. For each subject, the scanning session requires approximately 90 minutes and takes place on a different day than the MEG scanning session.

**Data processing**

*MRI data*

The fMRI data are processed using FreeSurfer [23], thanks to which, the brain is anatomically parcellated into 68 Regions of Interest (ROIs) or parcels using the automatic parcellation ('aparc') annotation. A neuroanatomical label is assigned to each ROI on a cortical surface model based on probabilistic information estimated from a manually labeled training set [23,24].

*MEG data*

We apply MaxFilter, an essential pre-processing tool for MEG data, in order to remove noise sources likely to originate from outside the sensor array. We then transform the MEG data using the temporally extended signal space separation method (tSSS) to remove strong interference caused by external and nearby sources. The tSSS-reconstructed MEG data are processed using MNE-Python [25,26], an open source Python library for the processing of EEG and MEG data. Next, the data are cleaned using Independent Component Analysis (ICA) to remove EOG and ECG artifacts. For each subject, the MEG recordings are co-registered to the anatomical fMRI preprocessed data. BEM, source, and forward solution for each run are then computed. Next, the MEG data are resampled at 500Hz, and notch filtered at 60Hz. Further filtering including low and high band filters at respectively 50Hz and 1Hz are applied as well. For each subject, the recording MEG runs are further concatenated in one single raw file. The precomputed forward solutions are averaged across runs and a covariance matrix is computed from the empty room MEG runs. The forward solution and the covariance matrix are used to compute an inverse solution. Using detected event ids corresponding to the stimuli presentation, we next proceed to the extraction of the events. The extracted events are epoched accordingly. From the previously computed inverse solution, the inverse operator is determined and applied to each of the epoched 0-back and 2-back conditions separately. The resting state MEG runs are processed similarly to the experimental runs without the event detection step. For each 0-back, 2-back, and resting state conditions, we compute the spectral coherence [27] to measure functional connectivity (FC) between MEG signals of ROIs or parcels $x$ and $y$ at a specific frequency band f as follows:

$$SC_{xy}(f) = \frac{|S_{xy}(f)|^2}{S_{xx}(f)S_{yy}(f)}$$

where $S_{xy}$ is the cross-spectrum between $x$ and $y$, and $S_{xx}$ and $S_{yy}$ are respectively the autospectra of $x$ and $y$. The connectivity matrix $SC(f)$ of size $68 \times 68$ where the entry $SC_{xy}(f)$ contains the value of the spectral coherence between the MEG signals of ROIs or parcels $x$ and $y$ at the frequency f. The connectivity matrices are computed at each and across *theta* (4 – 8Hz), *alpha* (8 – 15Hz), *beta* (15 – 35Hz), and *gamma* (35 – 120Hz) frequency bands. All data processing is performed using MNE-Python, an Open-source Python software [28].

**Network generation**

The computed connectivity matrices are adjacency symmetric matrices representing undirected weighted network, where the vertices are the 68 ROIs or brain parcels generated from the 'aparc' annotation and the edges are weighted by the magnitude of the spectral coherence. The adjacency matrices are then filtered to obtain the strongest edges in each brain network. While various studies [6,29–31] recommend different filtering techniques of the adjacency matrix, we decide to set an arbitrary threshold depending on each connectivity matrix. Using NetworkX [32], a python library for exploring complex networks, we generate binary functional brain connectivity networks from the filtered adjacency matrices. Each one of the graphs are exported in a graphml format for model estimation in R, an open-source environment for statistical computing [33].

**Assessing the small worldness of the connectivity networks**

Small world networks interposed between random and lattice networks. Like a regular lattice, they show high clustering and like regular random networks, they display low average path length. While the high clustering supports degeneracy and triangular integration, and may facilitate functional specialization, the low average path length facilitates efficient integration across the brain network. Since healthy brain networks have been proved to have small world organization [9], these two properties of small world networks have been used in clinical applications, particularly in the classification of brain disorders. [34,35]. To assess the small worldness of the generated functional connectivity networks, there remains the question regarding which clustering coefficient values should be considered high and which average path length values should be deemed as low. To address this question, Fornito et al. [36] propose a simple solution which consists in comparing the clustering and average path length values in each of the observed functional connectivity networks to comparable values computed in appropriately randomized control networks. Consequently, two indices which we adopt here, are defined:

- The normalized clustering coefficient $\gamma$ defined as:

$$\gamma = \frac{Cl}{\langle Cl_{rand} \rangle}$$

Where $Cl_{rand}$ is the average clustering coefficient computed over an ensemble of randomized surrogate network and $Cl$ is the average clustering coefficient of the observed network defined as:

$$Cl = \frac{1}{N} \sum_{i \in N} \frac{2t_i}{k_i(k_i - 1)}$$

Where $N$ is the number of nodes, $k_i$ is the degree of node $i$, and $t_i$ is the number of closed triangles attached to node $i$ in the observed network.

- The normalized measure of path length $\lambda$ defined as:

$$\lambda = \frac{L}{\langle L_{rand} \rangle}$$

Where $\langle L_{rand} \rangle$ is the mean of the average path length computed over an ensemble of randomized surrogate network, and $L$ is the observed average path length defined as:

$$L = \frac{1}{N(N-1)} \sum_{i,j \in N; i \neq j} d_{ij}$$

Where $d_{ij}$ is the distance of the shortest path, between nodes $i$ and $j$.

In a small world network therefore, one would expect $\lambda \sim 1$ and $\gamma > 1$.

Humphries et al. [37] proposed the ratio of $\gamma$ and $\lambda$ as a single scalar index to quantify the small-worldness of a network:

$$\sigma = \frac{\gamma}{\lambda}$$

A network with small world properties should be associated with a value of $\sigma$ greater than 1.

For each of the connectivity networks, we constructed an ensemble of 1,000 surrogate random networks using Monte-Carlo based simulations. We next compute respectively $\gamma$, $\lambda$, and $\sigma$ as defined above. Any network with a value of $\sigma$ greater than 1 is characterized as having small world properties. Since all our data have been recorded from "healthy individuals", we expect all the functional connectivity networks to display small world organization across all three conditions (0-back vs 2-back vs resting state).

**Statistical Group Analysis**

After the computation of the spectral connectivity in MNE-Python, the ROIs are exported in MNI coordinates in millimeters. The connectivity matrices are also exported as connectivity matrix files. Each matrix file contains the 68 lines by 68 columns of connectivity values. We then use the Network Based Statistic Toolbox (NBS) developed in Matlab by Zalesky et al. [38] to compare the brain networks between conditions. We used a non-parametric paired permutation t-test comparing the three conditions pairwise with a statistical significance level set at 0.05. The number of permutations is set at 100,000 for each comparison.

**Exponential Random Graph Model Estimation**

Given a network graph $G = (V, E)$, where $V$ is the set of vertices and $E$ is the set of edges, let the matrix $\mathbf{Y} = [Y_{ij}]$, be the random adjacency matrix of $G$. Each entry $Y_{ij}$ denotes a binary variable indicating the presence or absence of edge between two vertices $i$ and $j$. Since our brain connectivity network is an undirected network, $Y_{ij} = Y_{ji}$. Let's denote the matrix $\mathbf{y} = [y_{ij}]$ a particular realization of $\mathbf{Y}$. The general formulation of ERGM has the form [11]:

$$\mathbb{P}_\theta(\mathbf{Y} = \mathbf{y}) = \left(\frac{1}{\kappa(\theta)}\right) exp\left\{\sum_H \theta_H g_H(\mathbf{y})\right\}$$

Where $H$ is a configuration in $G$, $g_H(\mathbf{y}) = \prod_{y_{ij} \epsilon H} y_{ij}$, $\theta$ is a vector of parameter, and $\kappa(\theta)$ is a normalization constant defined as:

$$\kappa(\theta) = \sum_{y} exp\left\{\sum_{H} \theta_H g_H(\mathbf{y})\right\}$$

Several variants of ERGM have been proposed [39], here we rely on the temporal ERGM variant proposed by Leifeld et al. [40] which applied without any temporal dependencies corresponds to a pooled ERGM. We refer the reader to Leifeld et al. [40] for a detailed explanation of the model. Our main assumption justifying this choice is that different brain processes are involved in the 0-back, and 2-back memory tasks. Therefore, all changes in the functional connectivity brain networks under each condition are attributable to variation according to an underlying ERGM. Since the subjects are dependent from each other, the estimates of the pooled ERGM reflect the average effects across all the subjects' brain networks under a specific condition.

We model several organizational and functional mechanisms of the brain including functional segregation and functional integration [41,42]. Functional integration refers to distributed processes defining brain function and is measured in connectomics by the average path length (already defined above) or the global efficiency $E_g$ defined as:

$$E_g = \frac{1}{N(N-1)} \sum_{i,j \in N; i \neq j} \frac{1}{d_{ij}}$$

Functional segregation refers to the idea that all vertices in the brain network (or ROIs or brain parcels) will display divergent pattern of activity and hence be statistically independent. In connectomics, functional segregation is measured by the clustering coefficient (already defined above) and the local efficiency $E_l$ defined as:

$$E_l = \frac{1}{N} \sum_{i \in N} E_g(G_i)$$

Where $G_i$ is the subgraph formed by the vertices connected to $i$.

Model construction and estimation are computed using the statistical software R [33]. In the **btergm** R package that we used, functional integration and functional segregation are already respectively coded as the GWNSP (Geometrically Weighted Nonedgewise Shared Partner distribution) and the GWDSP (Geometrically Weighted Dyadwise Shared Partner distribution) ERGM terms [6]. We also model other ERGM terms including degree distribution, k-triangles (for transitivity) and k-stars (for highly connected vertices). We assess the Goodness-Of-Fit (GOF) of each model, simulating 1,000 networks from the estimated model and comparing them to the observed networks. The best model is selected based on the lowest Akaike Information Criterion (AIC) or the Bayesian Information Criterion (BIC) and the highest log likelihood.

The R packages **igraph** [43], **sna** [44] and **network** [45] are also used for the manipulation of the brain network graphs. All computations are performed in Rstudio-server setup on a 64 cores CPU server equipped with a 512GB RAM.

# RESULTS

In this section, we present pilot results based on a subset of nine subjects out of the 32 participants we collected neuroimaging data from. Likewise, only results on the 0-back and 2-back conditions are presented as the resting state data were yet to be processed at the pilot stage. Also, these pilot results were obtained from the connectivity matrices computed across all the frequency bands.

## Small-worldness Assessment

The results of the small-worldness assessment are presented in table 1. As we can see, across all subjects for the 0-back and the 2-back conditions, the functional connectivity brain network have values for $\gamma$ that are larger than one and values for $\alpha$ that are close to one. Consequently, the values for $\sigma$ are all larger than one. This is a strong evidence suggesting that all the functional connectivity brain networks have small-world properties.

Table 1. Small-worldness assessment of the brain networks based on 1000 randomized control surrogates

|  | 0-back | | | 2-back | | |
| --- | --- | --- | --- | --- | --- | --- |
| subjects | $\gamma$ | $\lambda$ | $\sigma$ | $\gamma$ | $\lambda$ | $\sigma$ |
| 1 | 4 | 1.107 | 3.613 | 4 | 1.104 | 3.623 |
| 2 | 3 | 1.156 | 2.595 | 3 | 1.157 | 2.593 |
| 3 | 4 | 1.167 | 3.428 | 4 | 1.226 | 3.263 |
| 4 | 5 | 1.16 | 4.31 | 5 | 1.161 | 4.307 |
| 5 | 4 | 1.142 | 3.503 | 4 | 1.18 | 3.39 |
| 6 | 4 | 1.214 | 3.295 | 4 | 1.205 | 3.32 |
| 7 | 3 | 1.113 | 2.695 | 3 | 1.12 | 2.679 |

**Group level comparison**

In the group level analysis, we see no significant difference (p>>0.1) between the connectivity values of the 0-back versus the 2-back conditions across all and for each of the frequency bands. This lack of significance is illustrated in figure 1 which displays the 300 strongest connections between the identified ROIs or brain parcels.

Figure 1. Connectivity plots of the 0-back (top) compared to the 2-back (bottom) memory tasks in subject 1.

**Exponential Random Graph Model**

Most of the model configurations we fit did not converge and/or degenerate. Table 2 presents the configurations of ERGMs we successfully fit at this pilot stage. At this stage, none of our model configurations containing the k-star or triangle ERGM terms was successful. All of them degenerated around 50 iterations and did not converge.

Table 2. Successful ERG model configurations

| Models  | edges | degree | GWDSP | GWNSP |
|---------|-------|--------|-------|-------|
| Null    | x     |        |       |       |
| Model 1 |       |        | x     |       |
| Model 2 | x     | x      | x     |       |
| Model 3 | x     | x      | x     | x     |

Table 3 presents the estimates of the ERGM configurations. We can see that the model estimates were all significantly higher than zero. However, for the null and model3, the confidence intervals of the estimates for the 0-back and 2-back conditions overlap meaning that those models failed to discriminate between both conditions. On the other hand, model1 and model2 discriminate between 0-back and 2-back conditions as the model estimates were significantly different than zero and their confidence intervals do not overlap. Overall, the ERGM model containing both the functional segregation and functional integration did not prove successful at discriminating between the 0-back and the 2-back conditions (see coefficient plot at Figure 2).

Given the low sample size, the AIC, BIC, and the log-likelihood were only computed for the null model. We could not efficiently compare the models according to those values. However, model3 containing the functional segregation and functional integration ERGM terms proves interesting as we believe an increase in the sample size would tremendously improve it at discriminating between 0-back and 2-back conditions.

Figure 4 shows the GOF plot of model 3 for both 0-back and 2-back conditions. The plain black line represents the feature distribution from the observed brain networks and the dashed black line is the feature distribution from the 1,000 simulated networks from model3. We expect both lines to overlap when the model captures the underlying ERGM process. As we can see in the GOF in figure 4, model3 captures well the underlying ERGM process for 0-back and 2-back. However, the simulated walktrap modularity distribution (in red) does not match well the observed one (in black).

**Conclusion**

In this report, we use a pooled variant of ERGM to capture differentially the underlying ERGM process involved in two nback memory tasks. Our models perform decently well given the significant model estimates. The low sample size of the brain networks is a tangible reason justifying the failure of most of the model configurations we attempted to fit. Consequently, we could not compare the model according to the AIC, BIC and the log-likelihood values. A larger sample size would enable a better model specification. The insignificant difference between the connectivity values of the 0-back and 2-back conditions at the group level comparison has been confirmed at the statistical modeling step. Nevertheless, the pilot results presented here are promising and would improve in robustness when all the remaining pre-processing will be completed and integrated to the analysis. While our results are not complete, they tend to support previous findings reported by De Vico Fallani et al. [5] that functional segregation and integration are sufficient to statistically reproduce the main properties of brain network.

It is worth noting that our connectivity networks were computed across all the frequency bands. Also, it would have been interesting to compare the resting state connectivity network pairwise with the ones of the 0-back and 2-back conditions. Unfortunately, those data were not pre-processed enough to be included in the analyses.

Finally, the connectivity values in this report are estimated by means of the spectral coherence which is known to suffer from possible volume conduction effects [46]. Other measures of connectivity such as Phase Lag Index (PLI), Phase-Locking Value (PLV), coherency, or the Imaginary coherence are potential alternatives worth considering. Although a binarizing threshold may influence the topology of the network, our thresholding procedure to filter the connectivity value and binarize the strongest edges has been based on the observation of the connectivity plot. A density based thresholding procedure has been proposed in [5] and proved to ensure a meaningful network.

Table 3.

| | 0back | | | | 2back | | | |
|---|---|---|---|---|---|---|---|---|
| | **Null** | **Model1** | **Model2** | **Model3** | **Null** | **Model1** | **Model2** | **Model3** |
| **edges** | -2.05 [-2.09; -2.00]* | | -0.85 [-0.94; -0.76]* | -2.63 [-2.83; -2.43]* | -2.06 [-2.10; -2.02]* | | 1.26 [1.21; 1.26]*** | -2.73 [-2.91; -2.55]* |
| **degree** | | | 1.64 [1.10; 2.18]* | 0.59 [0.09; 1.10]* | | | -4.22 [-4.39; -4.05]*** | 0.26 [-0.24; 0.75] |
| **GWDSP** | | -0.34 [-0.34; -0.33]* | -0.22 [-0.24; -0.20]* | 1.48 [1.29; 1.67]* | | -0.54 [-0.57; -0.52]* | -41.16 | 1.47 [1.31; 1.64]* |
| **GWNSP** | | | | -1.72 [-1.92; -1.53]* | | | | -1.71 [-1.87; -1.55]* |
| **AIC** | 145963.24 | | | | 245249.42 | | | |
| **BIC** | 145982.51 | | | | 245269.7 | | | |
| **Log Likelihood** | -72979.62 | | | | -122622.71 | | | |

***p < 0.001, **p < 0.01, *p < 0.05 (or 0 outside the confidence interval)

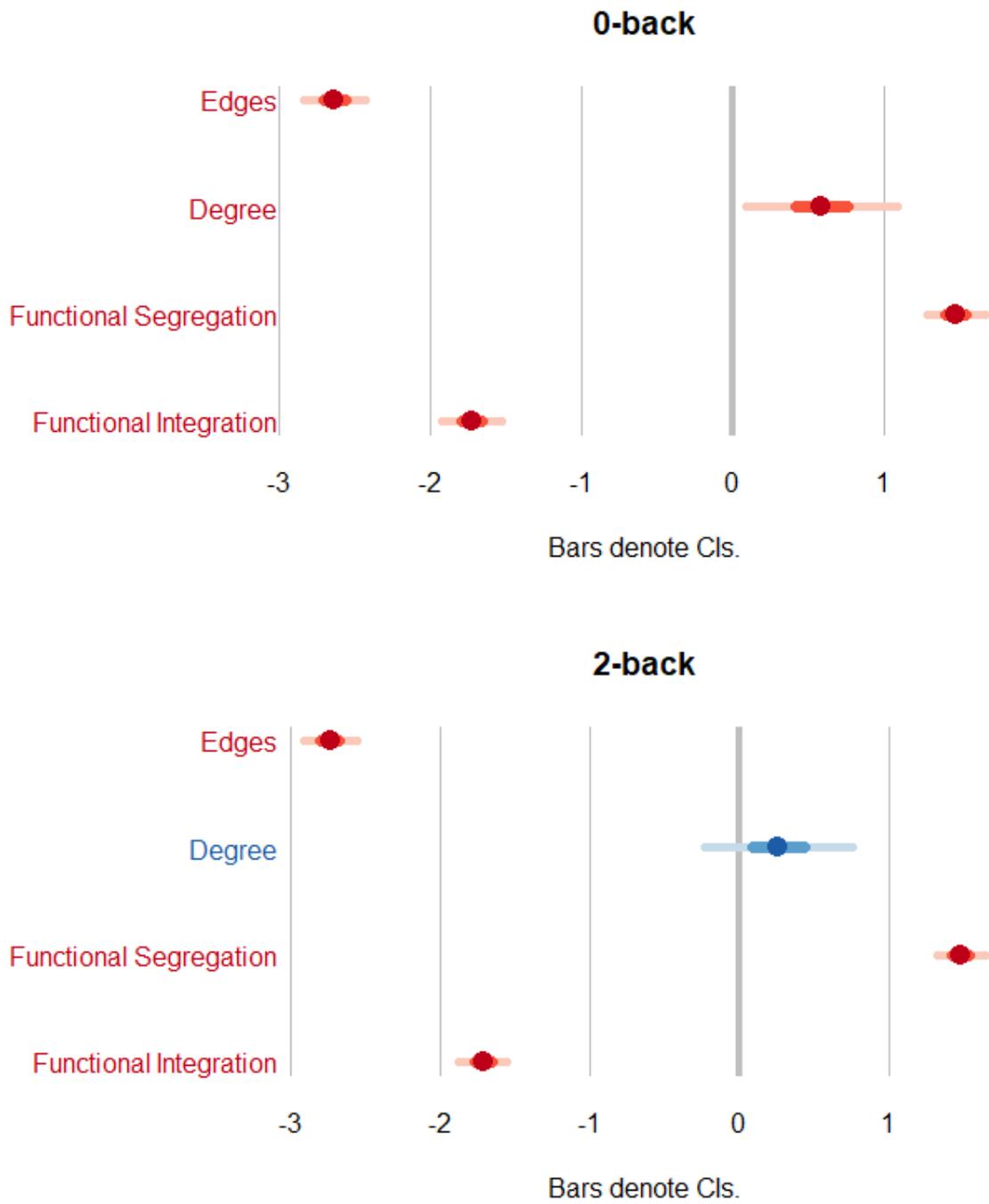

Figure 2. Coefficient plot of model3 comparing 0-back and 2-back conditions

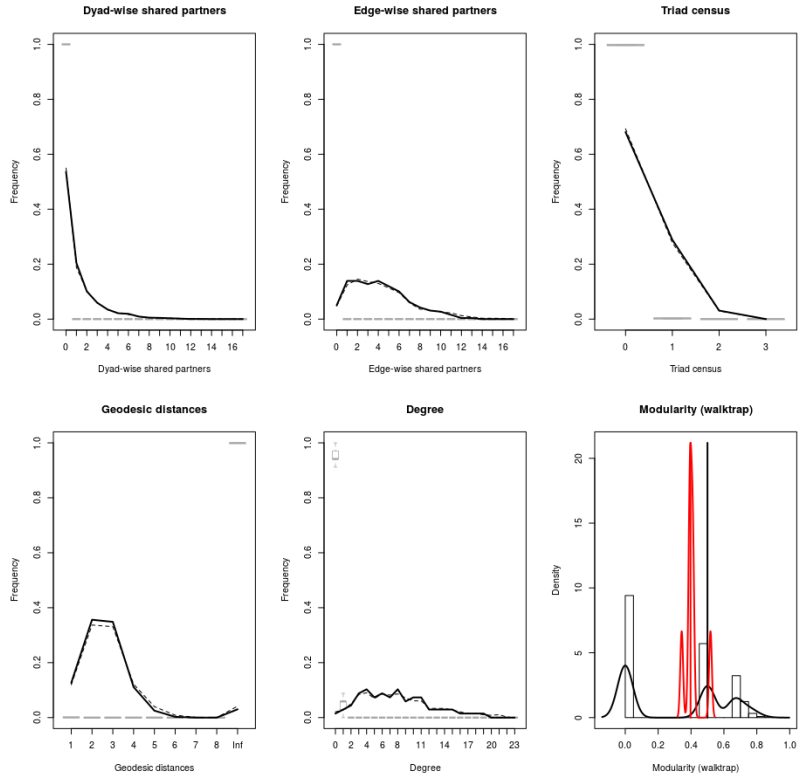 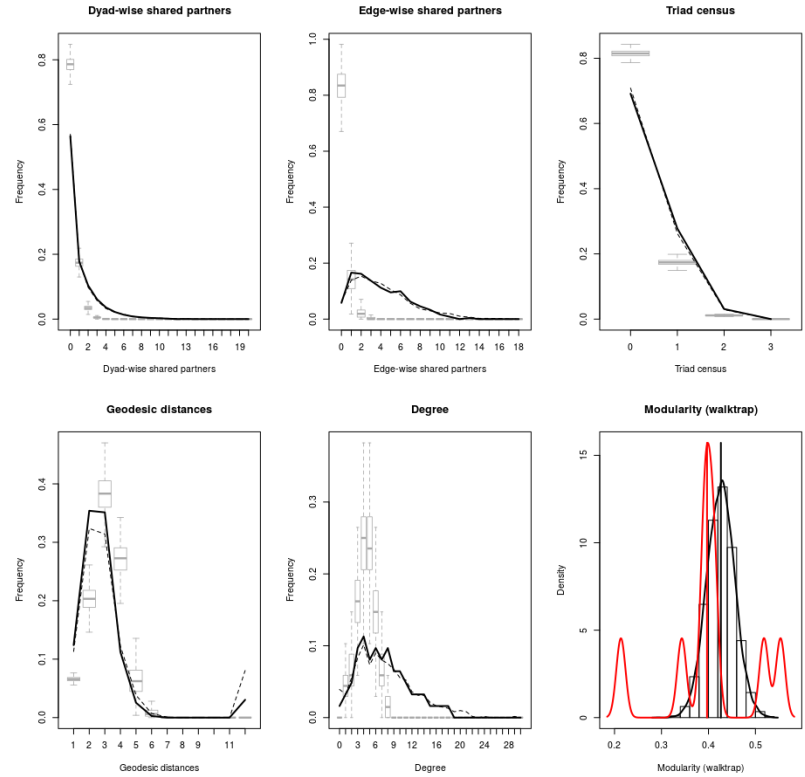

(a) Model3 GOF for 0-back condition

(b) Model3 GOF for 2-back condition

Figure 3. GOF of model3 comparing 0-back and 2-back conditions.